\documentclass[aps,prl,twocolumn,showpacs]{revtex4-1}

\usepackage{amsmath,amssymb}
\usepackage{bm}
\usepackage[dvips]{graphicx}

\def \G{\mathcal{G}}

\def \A{\mathcal{A}}

\def \w{\omega}

\def \Im{\textrm{Im }}

\begin{document}
\title{Self-consistent Green's functions calculation of the nucleon mean-free path}
\date{\today}

\author{Arnau Rios}
\affiliation{Department of Physics, Faculty of Engineering and Physical Sciences, University of Surrey, 
Guildford, Surrey GU2 7XH, United Kingdom}
\email[]{a.rios@surrey.ac.uk}

\author{Vittorio Som{\`a}}
\affiliation{CEA-Saclay, IRFU/Service de Physique Nucl{\'e}aire,
F-91191 Gif-sur-Yvette, France}

\begin{abstract}
The extension of Green's functions techniques to the complex energy plane provides access to fully dressed quasiparticle properties from a microscopic perspective. Using self-consistent ladder self-energies, we find both spectra and lifetimes of such quasiparticles in nuclear matter. With a consistent choice of the group velocity, the nucleon mean-free path can be computed. Our results indicate that, for energies above 50 MeV at densities close to saturation, a nucleon has a mean-free path of 4 to 5 femtometers.
\end{abstract}


\maketitle

The mean-free path, $\lambda$, of a nucleon in the nuclear medium is a basic transport coefficient, essential ingredient of several theoretical and experimental considerations. Cross section calculations within Glauber theory and transport simulations, for instance, rely directly or indirectly on in-medium mean-free paths in an energy range from above the Fermi surface to a few hundred MeVs \cite{Renberg1972,Yuan1989a,Glavanakov2004,Sinha1983,Sommer2011}. Below the Fermi energy, large values of the nucleon mean-free path provide a justification for the shell model \cite{Blatt1991}. At the Fermi surface itself, the existence of quasiparticles (qp) is validated by an infinite value of $\lambda$ \cite{Abrikosov1965}. 

Strong interactions between nucleons and the many-body correlations they induce play an essential role in nuclear systems \cite{Dickhoff2005}. Their existence precludes the application of mean-field or density functional techniques in the calculation of $\lambda$. As a matter of fact, the fully Pauli-blocked nature of these approximations leads to an infinite mean-free path below the Fermi surface. More sophisticated theoretical approaches have been used in the past
\cite{Negele1981,Fantoni1981a,Mahaux1983,Li1993}, but even then the hole mean-free path is not necessarily well defined \cite{Zuo1999,Sammarruca2008}. Moreover, calculations have been generally performed at energies arbitrarily close to the real axis and, as a consequence, \emph{ad hoc} nonlocality corrections have to be introduced, generally through the $k-$mass approximation \cite{Negele1981}.

Many-body Green's functions deal explicitly with time propagation of particles and holes in interacting fermionic systems, constituting a natural framework to compute qp properties \cite{Dickhoff2005,Economou2006}. 
Pauli principle and nucleon-nucleon (NN) correlations are fully taken into account. 
As we shall see in the following, the extension of these techniques to the complex energy domain provides a more consistent approach to compute qp quantities, without further approximations \cite{Fradkin1959,Negele1981,Economou2006}. Direct calculations on the complex energy plane have been performed in electronic systems since the early 1960's \cite{Engelsberg1963} and have been recently employed to describe microscopic excitations in solid state applications \cite{Eiguren2008,*Eiguren2009}. 
In this Letter, we devise such an extension for the case of nuclear matter and compute qp spectra and lifetimes, which eventually lead to an evaluation of the nucleon mean-free path from a fully microscopic standpoint. 

In the past, the lack of microscopic propagators in nuclear systems has hampered the calculation of transport coefficients using Green's function techniques, unlike other approaches \cite{Benhar2007a}. The recent implementation of the ladder approximation within the self-consistent Green's functions (SCGF) framework, however, gives access to the full off-shell energy and momentum dependence of one-body (1B) propagators. We refer the reader to Refs.~\cite{Frick2005,Soma2008} for details, but mention here that the method takes into account short-range and tensor correlations. Three-body forces (3BF) are included effectively via an average over a third, correlated nucleon \cite{Soma2008}. 

\begin{figure}[t!]
\begin{center}
       \includegraphics[width=0.9\linewidth]{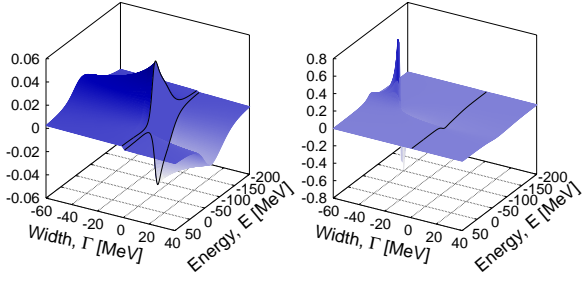}
       \caption{Imaginary part of the propagator in the complex energy plane, $z=E+i\Gamma$, at zero momentum for a CD-Bonn self-energy at $\rho=0.16$ fm$^{-3}$ and $T=5$ MeV. The left plot corresponds to the usual propagator, $\G(k,z)$, while the right plot represents its analytic continuation, $\tilde \G(k,z)$. Solid lines show the imaginary part of the propagator just above and below the real axis, $\pm \A(k,\w)/2$. }
       \label{fig:g_complex}
\end{center}
\end{figure}

Let us discuss the extension of the Green's functions formalism to the complex energy domain, before examining the calculation of qp properties.
The propagation of an excitation in nuclear matter is described by the retarded propagator,
$\G_R(k,t) \equiv \Theta(t) \left\langle \left\{ a(k,t), a^\dagger(k,0) \right\} \right\rangle$.
In a uniform system in thermal equilibrium, $\G_R$ only depends on the time difference, $t$, and the momentum modulus, $k$. Using the Lehmann representation, one finds the retarded propagator in energy space:
\begin{align}
	\G_R(k,\w)&=\int \frac{\textrm{d} \w'}{2\pi} \frac{\A(k,\w')}{\w_+-\w'}  \, ,
	\label{eq:gr_energy}
\end{align}
where $\w_+ \equiv \w+i\eta$ with $\eta \to 0$. All 1B operators can be built from the spectral function, $\A(k,\w)$. The previous expression suggests an extension of the propagator to the complex energy plane, replacing $\w_+$ by $z$:
\begin{align}
	\G(k,z)& \equiv \int \frac{\textrm{d} \w'}{2\pi} \frac{\A(k,\w')}{z-\w'}  \, .
	\label{eq:g_complex}
\end{align}
Such an extension exists, is unique, and gives rise to a complex variable function that is analytic off the real axis \cite{Baym1961}. Close to this axis by above (below), $\G(k,z)$ becomes the retarded (advanced) propagator, $\G(k,\w_\pm)=~\G_{R/A}(k,\w)$. 
An illustration of the complex energy dependence of the 1B propagator is given in the left plot of Fig.~\ref{fig:g_complex}. The discontinuity of $\G$ across the real axis is determined by the spectral function, $\textrm{Im}\left\{ \G(k,\w_-)-\G(k,\w_+) \right\}= \A(k,\w)$.

In the complex plane, the 1B propagator fulfills a Dyson equation
\begin{align}
\G(k,z)&= \frac{1}{z - \Sigma(k,z)} \, ,
\label{eq:dyson}
\end{align}
with a self-energy $\Sigma$ which is extended to complex energies in analogy with Eq.~(\ref{eq:g_complex}):
\begin{align}
\Sigma(k,z)&\equiv\int \frac{\textrm{d} \w}{2\pi} \frac{\gamma(k,\w)}{z-\w}  \, .
\label{eq:s_complex}
\end{align}
Our strategy will consist in computing $\gamma(k,\w)$ from full off-shell ladder self-energies and using Eqs.~(\ref{eq:s_complex}) and (\ref{eq:dyson}), to extend $\Sigma$ and $\G$ to the complex plane. 

Let us now illustrate how the mean-free path can be extracted from $\G(k,z)$. One computes $\lambda(k)$, from the nucleon inverse lifetime $\Gamma(k)$ and group velocity $v(k)$, via $\lambda(k) = v(k) / | \Gamma(k) |$. Such quantities appear in the asymptotic form of the retarded 1B propagator \cite{Dickhoff2005,Economou2006}:
\begin{align}
	\G_R(k,t) \xrightarrow[t >> \Gamma^{-1}]{}
	 -i \, \eta(k) e^{-i \varepsilon(k) t} e^{- | \Gamma(k) | t} \, .
	\label{eq:longtime}
\end{align}
While $\eta(k)$ represents the strength of the excitation, the qp spectrum $\epsilon(k)$ determines the oscillation frequency of the propagator. The spectrum is closely related to the group velocity
\begin{align}
	v(k)=\frac{\partial \varepsilon(k)}{\partial k}=\frac{k}{m^*(k)} \, ,
	\label{eq:groupvel}
\end{align}	
which is often studied in terms of an effective mass $m^*(k)$. 

It is natural to derive Eq.~(\ref{eq:longtime}) by Fourier transforming $\G_R$, under the assumption that $\G$ has a pole of order 1 in the lower half-plane \cite{Economou2006,Rios2011c}. The single pole description is particularly attractive due to its simplicity and its relation to Fermi liquid theory \cite{Abrikosov1965}. However, as we have seen, $\G$ is analytic for complex $z$ and, as a matter of fact, does not have a pole in the  complex plane. To find Eq.~(\ref{eq:longtime}), one actually trades the discontinuity of $\G$ across the real axis for a pole in the lower half-plane. The pole, however, is not associated to $\G$, but rather to its analytical continuation (ac), $\tilde \G$, which is continuous across the real axis but nonanalytic in the lower half-plane \cite{Rios2011c}. $\tilde \G$ is computed from the complex Dyson equation, Eq.~(\ref{eq:dyson}), with $\Sigma$ replaced by the ac of the self-energy, $\tilde \Sigma$:
\begin{align}
 \tilde \Sigma(k,z)  \equiv
\left\{
\begin{array}{c l}
  \Sigma(k,z) , &  \Im z > 0 \\
  \Sigma^*(k,z) , & \Im z \leq 0
\end{array}
\right. \, ,
\label{eq:g_cont}
\end{align}
a function that is analytic everywhere in the complex plane. 
The right plot in Fig.~\ref{fig:g_complex} shows the imaginary part of $\tilde \G(k=0,z)$ for a CD-Bonn SCGF self-energy. $\tilde \G$ is analytic across the real axis, but develops an isolated pole in the lower half-plane. The position of this pole is given by the complex equation:
\begin{align}
	z(k) = \frac{k^2}{2m} + \textrm{Re} \tilde \Sigma(k,z(k)) + i \Im \tilde \Sigma(k,z(k)) \, .
	\label{eq:fullpole}
\end{align}
The solution, $z(k) = \varepsilon(k) + i \Gamma(k)$, gives access to the fully dressed qp spectrum and inverse lifetime. 
As we have access to the ac of SCGF self-energies in the complex plane via Eq.~(\ref{eq:s_complex}), we are able to compute fully dressed spectra and lifetimes for different momenta, densities, and temperatures. Note that the solution to the previous equation need not be unique in the most general case. In our nuclear matter calculations, however, we have not found any signature of multiple solutions. 

\begin{figure}[t!]
\begin{center}
       \includegraphics[width=1\linewidth]{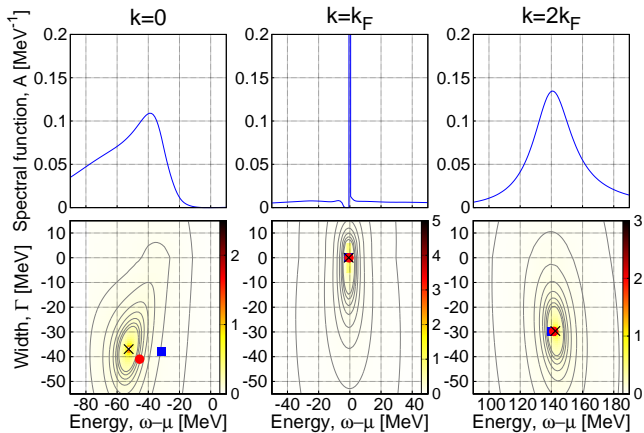}
       \caption{Upper panels: spectral function at $\rho=0.16$ fm$^{-3}$ and $T=0$ MeV for the CD-Bonn interaction. Lower panels: absolute value of $\tilde G$ in the same conditions. The fully dressed pole is indicated by a cross, while the circle (square) show the position of the first (second) renormalization quasiparticle.}
       \label{fig:g_contour}
\end{center}
\end{figure}

Previous calculations have relied on solving Eq.~(\ref{eq:fullpole}) using successive approximations for the complex energy dependence of $\tilde \Sigma$ \cite{Negele1981,Fantoni1981a,Mahaux1983}. At the lowest order, known as \emph{first renormalization} \cite{Eiguren2009}, one completely neglects the dependence on the imaginary part of $z$. This provides the usual definition of a qp: 
\begin{align}
	\varepsilon_1(k) &= \frac{k^2}{2m} + \textrm{Re} \tilde \Sigma(k,\varepsilon_1(k)) \, , \\
	\Gamma_1(k) &=  \Im \tilde \Sigma(k,\varepsilon_1(k)) \, ,
\end{align}
which usually coincides with the peak of the spectral function. A \emph{second renormalization} qp pole is obtained by expanding the self-energy around $z_1(k)$ to first order in the imaginary part of $z$,
\begin{align}
	\varepsilon_2(k) &= \varepsilon_1(k) - \Im \tilde \Sigma(k,\varepsilon_1(k)) 
	\, \Im \frac{1}{1-  \tilde\Sigma'( z_1 (k))} \, , \\
	\Gamma_2(k) &=  \Gamma_1(k) \,
	\textrm{Re} \frac{1}{1-  \tilde \Sigma'( z_1 (k))} \, .
\end{align}
In the context of nuclear physics, it has generally been assumed that the dependence of $\tilde \Sigma$ on the imaginary part of $z$ is soft and can be ignored in the previous derivatives \cite{Negele1981}. This gives rise to a slightly different qp pole:
\begin{align}
	\varepsilon_{2'}(k) &= \varepsilon_1(k) 
	\label{eq:spe_np} \\
	\Gamma_{2'}(k) &=  \Gamma_1(k) 
	\frac{1}{1-   \textrm{Re} \tilde \Sigma'( \varepsilon_1 (k))}  \, .
	\label{eq:spg_np} 
\end{align}
As we shall see, this approximation is well justified only above $k_F$.

In the following, we present our fully dressed results and compare them to previous approximations at $\rho=0.16$ fm$^{-3}$. The upper panels of Fig. \ref{fig:g_contour} show the SCGF spectral function, as a function of energy, for three different characteristic momenta ($k=0, k_F$, and $2k_F$). These have been obtained from a $T=0$ CD-Bonn self-energy \cite{Soma2008}. The lower panels give the absolute value of the analytically continued propagator. Contour levels unambiguously demonstrate the existence of a pole in $\tilde \G$. The location of the fully dressed pole is consistent with the numerical solution of Eq.~(\ref{eq:fullpole}), shown with a cross. Differences between this pole and the first or second renormalization properties are visible at $k=0$. At and above the Fermi surface, discrepancies disappear and the fully dressed pole coincides with first and second renormalizations. This points towards a very soft dependence of $\Sigma$ on the imaginary part of $z$ for $k \geq k_F$. Note that, at the Fermi surface, calculations yield a zero width, providing a verification of Fermi liquid theory from a self-consistent perspective \cite{Dickhoff2005}. 

\begin{figure}[t!]
\vspace{.15cm}
\begin{center}
       \includegraphics[width=0.85\linewidth]{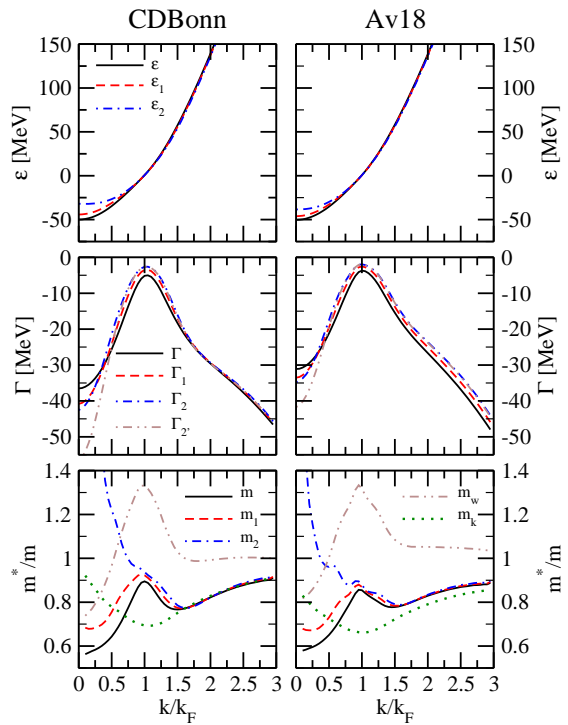}
       \caption{quasiparticle properties at $\rho=0.16$ fm$^{-3}$ and $T=5$ MeV for different approximations and two NN potentials: CD-Bonn (left) and Av18 (right panels). Upper panels correspond to the qp spectrum, central panels to inverse qp lifetimes, and lower panels to effective masses. The different approximations are explained in the text.}
       \label{fig:NNcomp}
\end{center}
\end{figure}

Nuclear many-body calculations are subject to uncertainties associated to the underlying NN interaction as well as to the approximation scheme itself. To assess them, we summarize in Fig.~\ref{fig:NNcomp} the results obtained with two different phase-shift equivalent potentials, the CD-Bonn \cite{Machleidt1995} and the Argonne $v_{18}$ (Av18) \cite{Wiringa1995} interactions, at $\rho=0.16$ fm$^{-3}$ and a finite, but rather small, temperature of $T=5$ MeV. The upper panels show the three approximations to qp spectra discussed earlier as a function of momentum. While above the Fermi surface the agreement between all approximations is good, below $k_F$ the fully dressed pole (solid line) is always more attractive than $\varepsilon_1$ (dashed line). In contrast, the second renormalization spectrum (dash-dotted line) is more repulsive. This indicates that successive renormalizations might not yield results closer to the fully dressed pole. The inverse qp lifetime, shown in the central panels, is bell shaped. Remarkably, below the Fermi surface the lifetime is finite. Close to $k_F$, its absolute value becomes small, but not zero due to thermal correlations \cite{Abrikosov1965}. Although not shown here, we have found that the effect of 3BF at this density is small in all the quantities shown \cite{Soma2008}. In contrast, many-body approximations other than GF's would yield rather different results. Within the Brueckner-Hartree-Fock approximation, states below the Fermi surface are completely blocked and $\Gamma=0$ for $k<k_F$ \cite{Zuo1999,Sammarruca2008}. 

To finish the calculation of the nucleon mean-free path,  a consistent determination of $\varepsilon$, $\Gamma$, and $v$, via Eq.~(\ref{eq:groupvel}), is needed. Let us, for instance, consider the nuclear physics renormalization, Eqs.~(\ref{eq:spe_np}) and (\ref{eq:spg_np}). The prefactor on the inverse lifetime is the inverse of the $\w-$mass. The group velocity involves the full effective mass, $\frac{m^*_1}{m}=\frac{m_\w}{m} \frac{m_k}{m}$. As a consequence, the mean-free path is only renormalized by the $k-$mass,
$\lambda_{2'}(k)~=~\frac{m}{m_k} \lambda_0$,
with respect to the uncorrected mean-free path,
$\lambda_{0}(k)~=~k/[2 m \, \Im \Sigma(k, \varepsilon_1(k)_+) ]$.
Thus, consistency in the spectrum and the lifetime are needed to obtain a nonlocality correction \cite{Negele1981,Fantoni1981a,Mahaux1983}. Similarly, the calculation of the self-consistent mean-free path relies on an effective mass computed from  the fully dressed spectrum rather than on $m_1^*$. 

\begin{figure}[t!]
\begin{center}
       \includegraphics[width=0.7\linewidth]{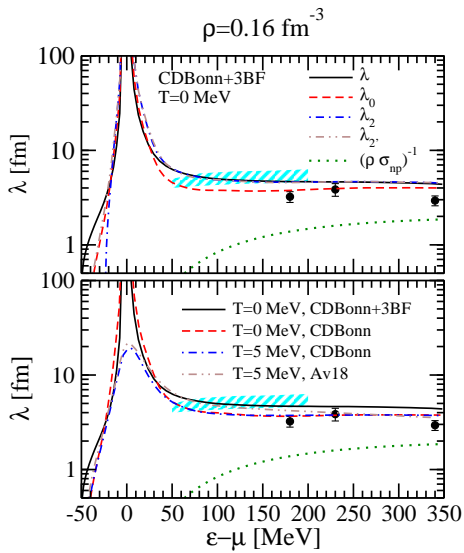}
       \caption{Mean-free path of a nucleon in nuclear matter as a function of energy. Upper panel: results obtained with a CD-Bonn+3BF self-energy at $T=0$ MeV. The different approximations are commented on the text. Lower panel: results obtained mean-free path from the fully dressed pole for different NN forces and two different temperatures. The shaded band and solid dots correspond to the experimental results of Refs.~\cite{Nadasen1981, Renberg1972}, respectively.}
       \label{fig:mfp}
\end{center}
\end{figure}

The lower panels of Fig.~\ref{fig:NNcomp} show the effective masses associated to the different approximations. Corresponding to a more attractive spectrum, the effective mass associated to the fully dressed pole (solid line) is lower than that of the usual qp approximation (dashed line). The latter is the product of the $\w-$ (double-dot dashed) and the $k-$masses (dotted). Our results confirm the well-known behaviors for these quantities: $m_\w$ peaks around $k_F$, while $m_k$ dips in this region. Note, however, that for the softer interaction (CD-Bonn), $\frac{m_\w}{m}$ becomes $1$ at lower momenta than for the harder force (Av18). For hole states, the effective mass associated with $\varepsilon_2$ shows a substantial increase as $k \to 0$, associated with the flattening of the spectrum in that region. 

A summary of final results is presented in Fig. \ref{fig:mfp}. The upper panel shows the mean-free path obtained with a fully realistic self-energy based on the CD-Bonn interaction supplemented with and Urbana-type 3BF \cite{Soma2008}. As expected, we find that the largest differences between approximations occur for hole energies, below $-20$ MeV, in a region where it is already relatively small. In contrast, above $50$ MeV, all approximations give similar results, except for $\lambda_0$ (dashed line), which is not corrected for nonlocality and thus should not be taken as a realistic prediction. $\lambda_{2'}$ (dash-dotted line) is only somewhat larger than $\lambda_0$ because of the small $m_k$ associated with the SCGF results. The kinetic theory prediction, $\lambda \sim (\rho \sigma_{np})^{-1}$ (dotted lines), is well below all quantum in-medium mean-free paths. The latter flatten at high energies, and remain constant, at a value of around $4-5$ fm. 

The lower panel of Fig.~\ref{fig:mfp} focuses on the NN interaction and temperature dependence of our results. The $T=0$ mean-free path with 3BF (solid line) is slightly larger than that obtained without 3BFs (dashed). The effect of temperature is relevant in an area of about $20$ MeV around the Fermi surface, where the mean-free path is finite, although still large. The fully correlated results agree with experimental estimates \cite{Nadasen1981, Renberg1972} and suggest that $\lambda \sim 4-5$ fm above $50$ MeV. The spread between different lines is an estimate of theoretical uncertainties, which amount to less than $1$ fm at those energies.

To summarize, we have devised a new method to obtain the mean-free path of a nucleon in the medium. The method involves the extension of Green's functions techniques into the complex plane. The pole of the propagator gives access to fully dressed qp properties. The renormalization induced by this procedure is relevant for hole properties. Our approach provides a validation for previously used approximations by taking into account the full dependence on the imaginary part of the energy. With all many-body corrections properly implemented, we obtain a mean-free path of around $4-5$ fm at saturation density and energies above $50$ MeV. Future work will systematically assess the density, temperature, and isospin asymmetry dependence of the mean-free path. 

\acknowledgments

This work has been supported by a Marie Curie Intra European Fellowship within the 7$^{th}$ Framework programme, STFC Grant No. ST/F012012, and by Espace de Structure Nucl\'eaire Th\'eorique (ESNT).

\bibliographystyle{apsrev}
\bibliography{biblio}

\end{document}